%% file: 00_main.tex
\documentclass[journal]{vgtc}   

\onlineid{0}

\vgtccategory{Research}

\vgtcpapertype{application/design study}

\title{Meandering Photo Lines:\\ Fluid Multiscale Exploration of Photographic Archives}

\author{%
  \authororcid{Mark-Jan Bludau}{0000-0001-6300-8833},
  \authororcid{Christian Tominski}{0000-0001-7704-355X}, and
  \authororcid{Marian Dörk}{0000-0002-3469-7841}
}

\authorfooter{
  \item
  	Mark-Jan Bludau (mark-jan.bludau@fh-potsdam.de) is with the University of Applied Sciences Potsdam and the University of Rostock (DE).
  \item
  	Christian Tominski is with the University of Rostock (DE).
  \item Marian Dörk is with the University of Applied Sciences Potsdam (DE).
}

\abstract{%
\input{sections/0_abstract}

}


\teaser{
  \centering
  \includegraphics[width=\linewidth, alt={Seven labeled screenshots of the Meandering Photo Lines interface. The top row shows four grid views transitioning from small multiples grid of abstract journey glyphs to round map-based and photographic tiles at different temporal scales. The bottom row shows the meandering path arrangement: first, its standard view with a timeline, journey path, globe follower, and image and keyword streams; then, a path with image-series piles, one of which is unfolded to reveal individual photographs. The final screenshot shows a full-screen black-and-white photograph of children studying the Torah in a workshop in Yemen, accompanied by its title, date, location, and keywords.}]{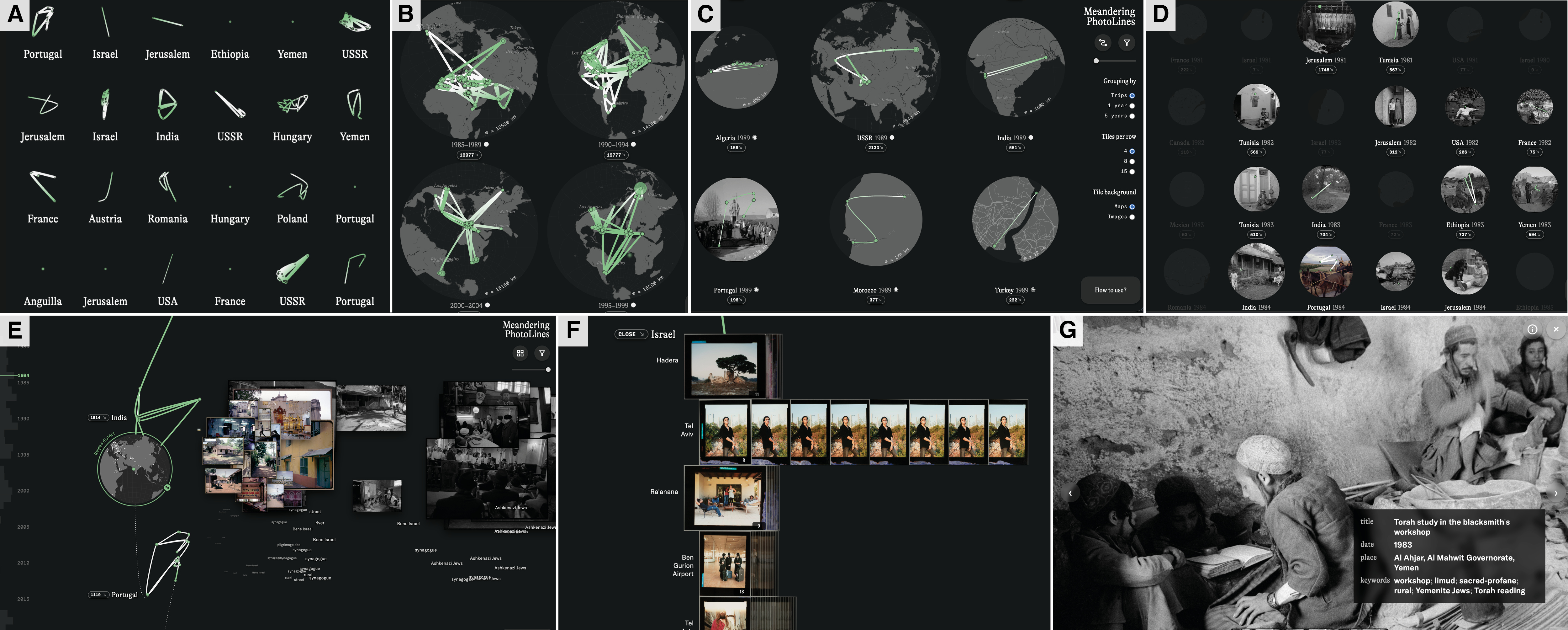}
  \caption{
  Meandering Photo Lines visualizes photographic archives as spatiotemporal journeys across scales, explored through a grid (A–D) and a path arrangement (E–G): (A)~Most zoomed-out grid, clustered by trips. (B)~Grid clustered by 5 years. (C)~Standard grid by trips with hover preview. (D)~Medium zoomed grid with a filter selected and image backgrounds. (E)~Standard path arrangement with timeline, journey, and image \& keyword stream. (F)~Unfolded journey showing image series. (G)~Detail view of a selected image.
  }
  \label{fig:teaser}
}

\graphicspath{{figs/}{figures/}{pictures/}{images/}{./}} 

\usepackage{tabu}                      
\usepackage{booktabs}                  
\usepackage{lipsum}                    
\usepackage{mwe}                       
\usepackage{wrapfig}
\usepackage{xcolor}

\usepackage{mathptmx}                  

\begin{document}


\firstsection{Introduction}

\maketitle

\input{sections/01_introduction}
\input{sections/02_relatedwork}

\input{sections/03_approach}
\input{sections/04_design}

\input{sections/05_usecases}
\input{sections/06_evaluation}

\input{sections/07_discussion}

\input{sections/08_conclusion}

\section*{Supplemental Materials}
\label{sec:supplemental_materials}
Supplemental materials include a demo video, high-resolution screenshots, and screenshot captions. A project website is available at \url{https://uclab.fh-potsdam.de/photolines}.

\acknowledgments{
We thank Frédéric Brenner and his team at the Frédéric Brenner Archive, all workshop and study participants, and Boris M{\"u}ller, Viktoria Br{\"u}ggemann, Jona Pomerance, Silvia Casavola, and Francesca Morini for their feedback. ChatGPT (OpenAI) was occasionally used to assist with debugging, code suggestions, and code improvements during prototype implementation. This work was funded by the German Research Foundation (DFG) through grants \href{https://gepris.dfg.de/project/510079995?lang=en}{510079995} and \href{https://gepris.dfg.de/project/514630063?lang=en}{514630063}.}

\bibliographystyle{abbrv-doi-hyperref-narrow}

\bibliography{00_bib}

\end{document}

%% file: sections/0_abstract.tex
We introduce a visualization that interweaves spatial and temporal structures of photo collections into explorable pathways. 
Existing photo browsing interfaces typically use space and time separately
for sorting, filtering, or mapping photos. What remains missing is an integrated visual form that represents the photographic journey itself. This is particularly relevant for the individual photographer's archive, where their unique lifeline is written into the collection, and it is this spatiotemporal trajectory that gives necessary context and meaning to the work. Inspired by the winding nature of rivers, fluid interaction, and un/foldable visualizations, \emph{Meandering Photo Lines} 
visualizes adjustable pathways through a photo collection across space, time, and topics. Two complementary views enable visual exploration at different scales: A grid-based arrangement of photo clusters offers a more analytical overview, while meandering spatiotemporal curves facilitate more serendipitous and reflective exploration. Animated transformations connect these two views seamlessly for multiscale navigation from spatiotemporal cluster signatures down to chronological photo series and individual photographs. We demonstrate our approach through two case studies: an archive of approximately 100,000 photographs documenting the Jewish diaspora life worldwide from 1978 to 2021 and a personal smartphone collection of around 20,000 pictures. Qualitative feedback from a think-aloud study indicates that our approach supports both analytical examination and curiosity-driven engagement.

%% file: sections/01_introduction.tex
The rise of digital photography has fundamentally changed how photographs are produced, stored, and accessed. While analog photography required deliberate decisions about each exposure, digital cameras---especially smartphones with automatic recording of timestamps and GPS coordinates---have made it easy to capture large numbers of images at minimal cost~\cite{hand2012ubiquitous,vanhouse2011personal}. This shift has both increased the volume of photographic material and enriched it with spatiotemporal metadata, which web platforms organize through temporal feeds, location data, and user-generated tags. Flickr, notably, named a user's feed a ``photostream,'' emphasizing the flow of photographs as a continuous, river-like sequence. At the same time, the growing size of photo collections has introduced challenges for organization and retrieval~\cite{whittaker2010easy}, primarily addressed through chronological sorting, folder structures, and search. In parallel, digitization efforts by museums and archives have made historically significant collections computationally accessible, leading to a growing body of work on visualizations for exploration~\cite{windhager2018visualization, whitelaw2015generous}.

\begin{figure}[t]
  \centering
  \includegraphics[width=\columnwidth, alt={Five narrow sections of a historical ribbon map arranged side by side, showing the meandering course of the Mississippi River.}]{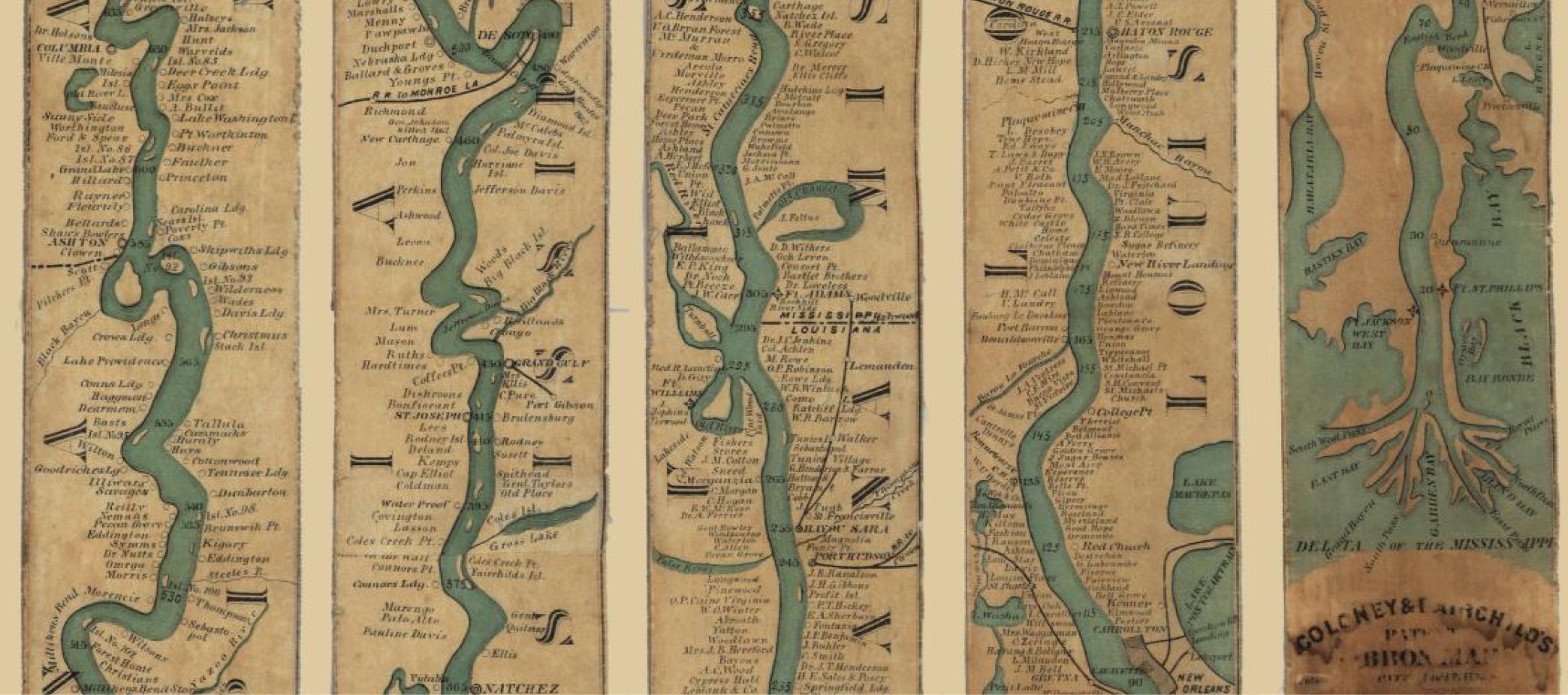}
  \caption{Ribbon map of the [Fa]ther of Waters (Coloney \& Fairchild, 1866). Source: Library of Congress, Geography and Map Division~\cite{coloney1866ribbonmap}. 
  }
  \label{fig:stripmap}
\end{figure}

Photo collection visualizations serve a range of tasks and contexts, from casual exploration in online interfaces~\cite{windhager2018visualization} and exhibitions~\cite{stefaner2018multiplicity}, to analytics tasks~\cite{van2017comparing}, browsing~\cite{kang2000visualization}, and reflection~\cite{kalnikaite2011saunter} of personal collections. Much of this work has addressed collections containing photos by multiple creators. Across these settings, interfaces are largely dominated by grids of thumbnails. Spatiotemporal information acts as secondary metadata for sorting, filtering, or coordinated views such as maps or timelines. The spatiotemporal structure of a collection receives no visual form of its own. This gap is particularly relevant for the individual photographer's archive, where the collection reflects a singular biography and a continuous creative process---a trajectory through space and time that provides crucial context for the work.

Drawing on fluid interaction~\cite{elmqvist2011fluid}, multiscale exploration~\cite{cakmak2021multiscale, jaenicke2015onclose}, and un/folding~\cite{bludau2025fluidly}, we introduce \textit{Meandering Photo Lines}, a visualization that interweaves spatial and temporal structures of photo collections into explorable pathways across multiple scales.  `Meandering' refers to a winding movement~\cite{oed_meander_n}, which reflects our interest in supporting casual, curiosity-driven navigation through a collection, while visually drawing on historic river strip maps (\autoref{fig:stripmap}). Our approach constructs journey paths from spatiotemporal clusters that can be navigated from distant overviews of movement signatures in a grid down to chronological photo series and individual photographs, connected through animated transitions and on-demand unfolding (\autoref{fig:teaser}).

We contribute (1)~a visualization technique for fluid multiscale exploration of spatiotemporal photo collections, and (2)~lessons from two case studies of different scale and type---a professional photographer's archive of approximately 100,000 digitized analog photographs and a personal smartphone collection of around 20,000 pictures---accompanied by qualitative feedback from a think-aloud study.

%% file: sections/02_relatedwork.tex
\section{Related Work}\label{relatedwork}

This research builds on prior work in exploratory visualizations of photo collections, as well as spatiotemporal and path-based visualizations. 

\subsection{Photo Collection Visualization}
Unlike abstract data, 
photo collections carry aesthetic quality through visual form and cultural meaning through depicted content.
Visualization design should balance abstract metadata 
with visual content and support distant reading of entire collections and detailed views of individual artifacts~\cite{windhager2018visualization}. However, traditional photo browsers largely rely on thumbnail lists and faceted search, which can limit their ability to convey context, scale, and cultural significance of a collection~\cite{whitelaw2015generous, windhager2018visualization}. 

Many photo browsers enable visual querying of metadata~\cite{kang2000visualization, wu2009photoscope, giannella2021fotovis}. Systems such as \textit{Photogrammar}~\cite{arnold2020visualizing} and \textit{KD-Photomap}~\cite{peca2011kd} combine thumbnail lists for browsing with coordinated maps and timelines for filtering.
While coordinated views may support analytical tasks, they oftentimes demand large screens to display all relevant information simultaneously and tend to have a more technical appearance.

Particle-like visualizations arrange small thumbnails into meaningful overviews, for example, based on color~\cite{hochman2013zooming}, genre~\cite{bludau2021relational}, or similarity~\cite{pixplot2017}. 
\textit{MyLifeBits} arranged photos and query results in or as timelines~\cite{gemmell2002mylifebits}, while later focus+context approaches used spiral-formed timelines organized in events~\cite{hilliges2007photohelix}, or zoomable and filterable image timeline layouts~\cite{glinka2017past}.  
Map-based approaches reach from labeling-algorithm-based thumbnail placements on maps~\cite{viana2007photomap} to more elaborate filter-highlight interactions~\cite{dork2012fluid} and force-driven inset labeling strategies~\cite{lekschas2019pattern}.
These approaches typically treat time and space as separate rather than combining them into one integrated representation.

The visual content of photographs can be transformed into reduced summarizations, for example, through slicing~\cite{lay2016slicing}, blending~\cite{viegas2007artistic}, collaging ~\cite{wallick2007automatic, yu2022softcollage}, or abstracted representations~\cite{viegas2009flickrflow}. Large collections can be aggregated into hierarchical layers arranged as treemaps~\cite{bederson2001photomesa}, zoomable grid layouts~\cite{frey2022optimizing}, piles~\cite{baur2008flux, lekschas2020generic}, collages~\cite{wallick2007automatic}, or origami structures~\cite{hsu2009phorigami}. High-level views can be unfolded on demand via semantic zooming, filtering, or selection~\cite{bludau2025fluidly}, 
while representative previews allow viewers to see more images~\cite{holmquist1998hierarchical}.
Hierarchical approaches often use algorithms to cluster semantically related or visually similar images~\cite{cooper2005temporal}. 
Piling~\cite{lekschas2020generic} and unfolding~\cite{bludau2025fluidly} techniques are particularly relevant for our work, as they provide mechanisms to foreshadow or progressively reveal underlying content within aggregated structures.

\subsection{Spatiotemporal Visualization}
Many historical examples combine temporal and geospatial data in map-related visualizations. A well-known example is Charles Minard's 1869 map of Napoleon's march to Moscow, combining the route, the army's diminishing size, and the freezing temperatures of the retreat~\cite{tufte1983visual}. Similarly, strip maps (\autoref{fig:stripmap}) have historically been used to represent paths, routes, or river courses in linear order
~\cite{tufte1991envisioning}. Related principles are used in transportation and route planning applications, where geospatial paths are flattened into simplified route diagrams to facilitate navigation~\cite{agrawala2001rendering}. 
Notably, these approaches straighten routes into simplified vertical paths, abstracting geographic orientation in favor of readability.

Beyond route-based examples, spatiotemporal visualizations range from timestamp-based representations and image series to space-time cubes and real-time 3D scene rendering~\cite{zhong2012spatiotemporal}, as well as trajectory visualization through the lens of geographic and abstract space, distinguishing spatial, temporal, and attribute-based representations~\cite{he2019variable}. 
While timelines can function as interlinked interactive interface elements~\cite{kraak2005timelines},
a recurring challenge is to combine spatial and temporal dimensions without introducing visual clutter as data scale increases.

Several projects have explored spatiotemporal visualizations as facilitators for reflective exploration of personal movement or shared reminiscing~\cite{meier2018data, thudt2015visual}. 
Particularly related to our work is \textit{Visits}~\cite{thudt2013visits}. It clusters personal GPS movement data into ``stays'' based on locations and time, and visualizes these stays as small miniature maps in combination with Flickr photo data. Similarly, \textit{Shifted Maps}~\cite{otten2018shifted} depicts personal movement data as multiple small maps connected in a network structure on a base map, highlighting relevant places in smaller map overlays. Both approaches define individual zoom levels and circular map cutouts based on data clusters. 
Our approach also employs such spatiotemporal clustering strategies, but additionally aims for integrating photographic material directly with the photographic journey, which the existing works do not offer.

Mayr and Windhager argue that coherent integrated views with seamless transitions reduce cognitive effort compared to coordinated multiple views, as they better support the integration of spatial and temporal information~\cite{mayr2018once}. Building on such integrated views, recent work explores narrative pathways through space-time cubes~\cite{eccles2008stories, mayr2018once} and a design space for spatiotemporal data stories~\cite{mayer2023characterization}, using techniques such as camera movements and zooms~\cite{li2023geocamera}, animated paths~\cite{nyt2013_russia_left_behind}, map morphing between related views~\cite{reilly2004map}, and adaptive composite map projections~\cite{jenny2012adaptive} to communicate journeys through space and time.

\subsection{Visualizations with Paths and Curves}
Temporal and geospatial sequences of events are commonly visualized as paths, which have a long tradition in cartographic and narrative contexts. From a humanities perspective, narrative structures in photo archives have been described as paths organized by attributes such as time, space, characters, and events~\cite{zeng2022storified}. 

Several visualization techniques make use of curved paths to represent sequential or evolving data. Map-based data stories often use animated paths that either grow as highlighted overlays on pre-drawn routes or are incrementally revealed during scrolling interaction, guiding readers along a spatiotemporal storyline~\cite{nyt2013_russia_left_behind}. 
\textit{TimeSeriesPaths}~\cite{Bernard12TimeSeriesPaths} and
\textit{Time Curves}~\cite{bach2015time} position data points sequentially on a timeline and spatially fold the path according to similarity, bringing related points closer together and creating unique new structures. Time Curves can be gradually transformed into a common timeline through a slider. Similarly, Chen et al. generate similarity-based curves for multiscale document exploration with focus+context selection and scale sliders~\cite{chen2012sequential}.

Paths have also been used for interaction. For example, \textit{DimpVis} supports temporal navigation through direct 
dragging of data elements along their temporal path in a visualization~\cite{kondo2014dimpvis}.
With \textit{Transmogrification}~\cite{brosz2013transmogrification}, drawing a path on a visualization or map creates a path-based shape that can be morphed into a new target shape, for example, to straighten routes into horizontal layouts. Similarly, \textit{Tied in Knots}~\cite{elli2020tied} interactively unfolds curved paths that visualize textual testimony on sexual harassment in academia into readable horizontal text fragments.

Paths and curves can also be used as visual signatures of data subsets. Reduced to small glyphs or fingerprints within small-multiples representations, visual signatures enable quick comparison across elements and have been applied in many visualization contexts~\cite{tufte1983visual, fuchs2013evaluation, hlawatsch2011flow, borgo2013glyph}, as well as in more artistic displays~\cite{hack2015sortedcities, lupi2016dear}.
Bach et al. generalized the design space of curves for data visualization in a framework describing visualization pipelines, encoding possibilities, and ways of embedding curves in visual representations~\cite{bach2018ways}.
Despite their versatility in terms of encoding, coherence, and comparability, curves and paths have not been used to represent spatiotemporal structures of photographic archives.

%% file: sections/03_approach.tex
\section{Approaching Photographic Journeys}\label{approach}

Prior visualization research offers various ways to explore image collections.
Spatiotemporal clustering, path-based sequential representations, and fluid interaction mechanisms offer relevant building blocks.
Across these approaches, however, space and time typically function as separate metadata dimensions---used for sorting, filtering, or coordinating views---rather than as an integrated visual form that reflects how a collection came into being. This is particularly limiting for an individual photographer's archive, where the collection is not an aggregation of images by different creators but the record of a continuous movement through space and time, shaped by decisions about where to go, what to photograph, and when to return. This trajectory provides context for understanding both individual images and the collection as a whole.

Representing this trajectory as a visual structure requires a tighter coupling of the photographic material with its spatiotemporal context, and representations that can operate across scales---from decades-long movement patterns down to individual photo sessions. While spatiotemporal clustering~\cite{thudt2013visits, otten2018shifted}, sequential data curves~\cite{bach2015time}, and multiscale unfolding~\cite{bludau2025fluidly} address parts of this challenge, they have not been brought together around the question of how a photographer's journey can become an explorable visual structure of a photo archive. Answering this question requires close engagement with the specifics of photographic archives---their data structures, cultural contexts, and modes of use---alongside concept development, data exploration, and prototyping in close exchange with archivists and domain experts.

\subsection{Design Methodology}

Our work follows an iterative development and reflection process with domain experts, similar to a design study~\cite{sedlmair2012design}.
The process was initiated through a collaboration with a photo archive on the Jewish diaspora by the professional photographer \textit{Frédéric Brenner} (see \autoref{casestudies}), aiming to provide a platform for visual access to the oeuvre. The visualization concept and archive cataloging evolved concurrently, with emerging data structures and design patterns informing each other.

\begin{figure}[t]
  \centering
  \includegraphics[width=\columnwidth, alt={Six labeled workshop collages arranged in two rows. They combine printed photographs, maps, handwritten annotations, and colored lines or strings to organize images as stacks, grids, sequences, timelines, and spatial arrangements.}]{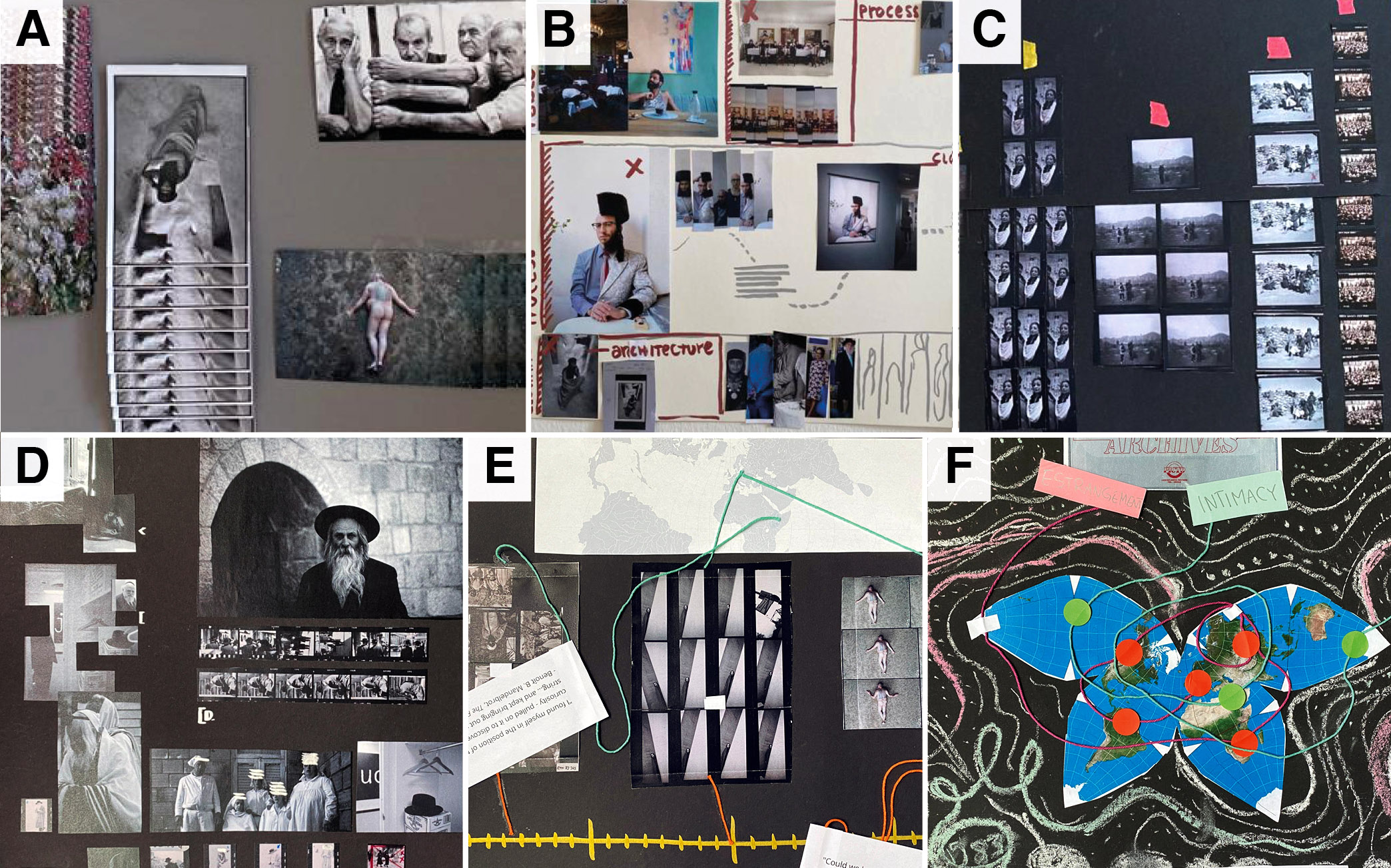}
  \caption{Selected collages from two co-design workshops.}
  \label{fig:collages}
\end{figure}

Throughout the collaboration, we worked with project partners and domain experts from photography, curation, archiving, and Jewish studies, as well as external participants with a general interest in the topic. To collaboratively identify promising pathways into photographic archives, we conducted two co-design workshops~\cite{chen2014exploring, bruggemann_2025_laying} with 10--20 participants each. In these sessions, participants created and discussed collages of the photographic material to identify potential entry points for exploration and visualization (\autoref{fig:collages}). The workshops were complemented by iterative sketching, experimental data explorations, and smaller prototypes, which were all discussed with project partners~\cite{doerk2020co-designing}.

Although the design process was embedded in a specific project collaboration, many of the questions it raised---about how to represent photographic processes, how to balance overview and detail, how to integrate spatial and temporal structures with the visual material---are relevant for the visualization of photographic archives more broadly.

\subsection{Conceptual Considerations}

Across both workshops and the iterative prototyping, several themes emerged that informed the direction of our work. We focus here on the spatiotemporal dimensions and the photographic journey.

\paragraph{Primacy of photographs.} Workshop participants consistently foregrounded the visual material itself, treating maps and timelines as supporting structures rather than primary modes of access:
\textit{``What I think is also interesting is that the photographs are kind of at the interface between time and space and drawing connections between them.''}
This confirmed the relevance of an approach where photographs remain visually present rather than being subordinate to abstract metadata.

\paragraph{Spatiotemporal dynamics and patterns.} Participants expressed interest in how temporal patterns could be visualized in relation to spatial context:
\textit{``[…] laying out the temporal dynamics over a map or over something, some other kind of display could be really fruitful, especially with encountering the different keywords and concepts.''}
A participant created a collage focusing solely on photos featuring hats in various settings (\autoref{fig:collages}D), indicating an interest in content-based filtering capabilities in combination with spatiotemporal comparison.
At the same time, partial, situated views were noted as preferable to comprehensive map displays. 
Due to the archive’s global dispersion, feedback on map-based prototypes indicated interest in non-Western centered projections to challenge conventional geographic framings.

\paragraph{Photographic processes.} A recurring theme in the collages was the serial character of photography (\autoref{fig:collages}A--C). Participants noted that many published images are part of larger photographic processes, including multiple shots of the same subject or scene as well as test shots across times and locations. Participants identified these serial structures as relevant for understanding the photographer's creative and selection processes:
\textit{``I mean the process part also tells a story from the image to the different options during creation and then how it is exhibited.''}
Furthermore, the biography of the photographer himself was seen as an overarching process:
\textit{``You come from the point of the photographer, look at kind of the bio of the photographer, what travels he had made, and […] dive through this into the story.''}

\paragraph{Gentle guidance.} Participants also discussed the tension between open exploration and the risk of disorientation in large collections:
\textit{``But at the same time, it's the question again that people don't get lost. [\ldots] We would need to also pull people along some path in order to not to have this reaction.''}
This points toward the potential of pathways through the collection to provide structure without imposing rigid navigation. Questions regarding the quantity of photographs and modes of guidance also arose from discussions of early prototypes, particularly around the contrast between immersion in the photographic journey and the presentation of individual well-known photographs.

\smallskip

Taken together, these initial observations and general aspirations suggested an approach that interweaves the photographic material with its spatiotemporal context, represents the serial nature of photography, and constructs navigable pathways to support both orientation and open-ended exploration. They also indicated the need for multiple scales of engagement---from overviews of movement patterns across decades and continents down to individual photographic situations.

\subsection{Tasks and Usage Contexts}
The workshops and discussions also helped to identify relevant tasks and audiences.
We derive four core tasks: 
(1) \textit{exploratory browsing} through overview generation and guided pathways, 
(2) \textit{close inspection} of individual photographs,
(3) \textit{spatiotemporal comparison} of patterns and content,
and (4) \textit{filtering} by topic. 
These tasks relate to existing task taxonomies, including Shneiderman’s~\cite{shneiderman2003eyes} ``Overview first, zoom and filter, then details-on-demand’’ and Windhager et al.'s~\cite{windhager2018visualization} elementary and synoptic cultural-heritage tasks (e.g., comparing patterns through exploratory analysis, or locating photographs by time or place).

Cultural collection interfaces typically address both casual and expert users~\cite{windhager2018visualization}, and our archive partners emphasized public accessibility, researcher-oriented capabilities, and potential exhibition use alike. Our approach therefore targets web-based exploration across display sizes---from handheld devices to large installations---requiring no expert knowledge to enter, while still supporting research-oriented tasks such as spatiotemporal comparison and qualitative process analysis.

\subsection{Meandering}

To provide a conceptual frame for the ambitions and tasks that emerged from the design process, we turn to the notion of \textit{meandering}. The verb to meander denotes \textit{``to follow a winding course''} or \textit{``to wander aimlessly''}~\cite{oed_meander_v}, while the noun meander may refer to \textit{``a circuitous journey or movement''}~\cite{oed_meander_n}. This double meaning---a spatial form and a mode of movement---corresponds to our interest in representing photographic journeys as winding paths while supporting open, curiosity-driven navigation through a collection.

Meandering relates to the photographic journeys of photographers themselves, which consist of spontaneous and planned photographs, sequences of images taken at the same place and time, and recurring themes across space and time. Revisiting such journeys through interactive experiences may take the form of open-ended exploration or more task-oriented navigation. In this sense, meandering also describes a mode of engaging with photo collections: akin to the movements of an information flaneur~\cite{dork2011information}, experiencing a collection without a specific goal, guided by curiosity and serendipity.

As a visual reference, we draw from historic river strip maps (\autoref{fig:stripmap}), which represent the course of rivers in vertical strips. These maps flatten a geospatial path into a readable sequence while preserving local geographic character---a principle that is relevant for constructing journey representations that maintain the spatial orientation of visited places while connecting them into a continuous path.

Translated into a visualization concept, meandering suggests that the movements of a photographer are represented as visually connected curves, inviting viewers to follow the journey at different scales. At the core of such an approach would be the construction of spatiotemporal paths from discrete space-time clusters, each producing an individual journey signature that preserves temporal sequence within geographic context. Connecting these signatures would form a representation of the whole journey that can be explored at different levels of detail.

\subsection{Design Goals}\label{approach:goals}

Based on the related work, the insights from the co-design process, and the conceptual framing of meandering, we formulate four design goals to orient the subsequent design and development.

\paragraph{\textbf{DG1: Place photographs in spatiotemporal context.}}
Unlike abstract data types, the visual appearance of photographs is central to their exploration. In humanities research, it has been described as beneficial to reduce the semantic distance between an artifact and its representation, for example through temporal and geospatial context and representations close to the original form~\cite{lamqaddam2020introducing}. We aim for a tighter coupling of the spatial, temporal, and visual dimensions of a photo collection to support contextualized exploration.

\paragraph{\textbf{DG2: Enable fluid multiscale exploration.}}
Visualizations of cultural collections offer the opportunity to combine distant overviews for pattern recognition with close inspection of individual artifacts~\cite{windhager2018visualization}. Increasing collection sizes require interaction strategies to navigate inherent hierarchical structures and to represent large amounts of material within limited space~\cite{bludau2025fluidly}. Visualizations often employ fluid interaction mechanisms, including animated transitions and direct interactions, to support coherent exploration across scales~\cite{elmqvist2011fluid, bludau2025fluidly}. Navigation across multiple scales should be supported, with smooth transitions between views that provide both overviews and spatiotemporal details.

\paragraph{\textbf{DG3: Support open-ended and task-oriented use.}}
Researchers often approach a collection with specific tasks and questions, requiring structured examination and analytical sensemaking. Conversely, casual users and exhibition visitors enter without a predetermined goal, guided by general curiosity, where low cognitive effort and intuitive interaction can promote serendipitous discoveries~\cite{mayr2016visualization, windhager2018visualization}. The visualization should accommodate both: goal-directed analysis of spatiotemporal patterns and thematic structures as well as open-ended, curiosity-driven meandering along the photographic journeys.

\paragraph{\textbf{DG4: Support a range of display contexts.}}
Personal photo collections are typically accessed on smartphones and tablets, while professional archives and exhibition installations may use desktop screens, kiosks, or projections. Rather than optimizing for a single display context, the visualization should follow interaction strategies that work across this spectrum---from handheld devices~\cite{horak2021responsive} to large public displays~\cite{Belkacem24LHRDSurvey}---so that the same spatiotemporal journey can be explored in personal reflection as well as in shared exhibition environments.

\smallskip

%% file: sections/04_design.tex
\section{Meandering Photo Lines}\label{meanderingphotolines} 

Our approach centers on the creation of continuous journey paths through photographic archives. Such an archive is a temporally sorted sequence of photographs tagged with timestamps, locations, and keywords, which we call the \textbf{photographic journey}. \textit{Meandering Photo Lines}\footnote{Project website: \texttt{https://uclab.fh-potsdam.de/photolines}} consists of several 
components that process and represent the spatial, temporal, and thematic aspects of such photographic journeys.

A fundamental computational step is to subdivide a journey into spatiotemporal clusters that capture relevant \textbf{journey fragments}, i.e., groups of photographs taken in close spatial and temporal proximity.

The core visual components are miniature maps, fragment paths, the global journey path, and the photographs themselves. \textbf{Miniature maps} represent the spatial context of journey fragments (DG1) and are overlaid with \textbf{fragment paths} that represent the spatiotemporal sequence of where and when photos were taken within a journey fragment. As such, fragment paths serve as visual signatures of journey fragments. The global \textbf{journey path} connects all journey fragments into one path. 

The journey path can be displayed in two distinct arrangements: (1) the \textbf{meandering path} and (2) the \textbf{aligned grid}. Both show the same sequence of journey fragments and are fluidly bridged through animated transitions (DG2). While the aligned grid primarily serves as an overview for structured, task-oriented exploration, the meandering path aims to support serendipitous discoveries along a journey (DG3). Either arrangement can be used as entry point to the interface, and both can be adapted through adjustable clustering scales and keyword filters, enabling the construction of alternative journeys at different spatiotemporal granularity or focused on specific topics. The interface is designed with responsiveness~\cite{horak2021responsive} in mind, following strategies such as vertical scrolling, grid-based layouts, and scalable sizing, which support adaptation across display contexts, from smartphones to large interactive installations (DG4).

\subsection{Constructing Journey Paths}

\begin{figure}[t]
  \centering
  \includegraphics[width= 0.86\columnwidth, alt={Diagram showing how a photo collection is grouped into journey fragments based on photo series, spatial, and temporal information. Each circular journey fragment consists of visited places represented as dots and a path connecting these places. Multiple fragments form a full journey, which is displayed either as a continuous meandering path of vertically arranged and connected fragments or as a grid of circular fragments.}]{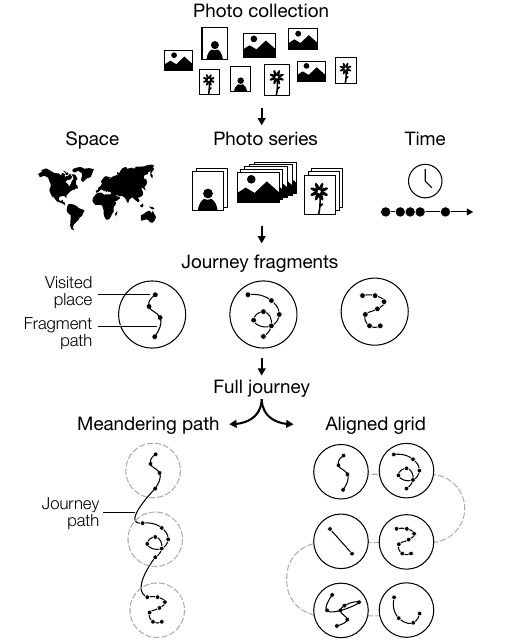}
  \caption{Construction of photographic journeys via spatiotemporal journey fragments and their display as meandering path or aligned grid.}
  \label{fig:concept}
\end{figure}

First, we define a hierarchy of journey units formed through flexible spatiotemporal clusters of photographs. Similar to \textit{Visits}~\cite{thudt2013visits}, these units are constructed across multiple levels of granularity (DG2), from individual photos to complete journeys (\autoref{fig:concept}).

Starting from \textbf{(1)~individual photos} as the smallest unit, each associated with temporal and spatial metadata, photos are grouped into \textbf{(2)~photo series} based on their serial character, representing variations of the same motif, scene, and setting (i.e., they share the same place and time). These series are then grouped into \textbf{(3)~visited places}, based on temporal sequence and location, where a new visited place is created whenever the location changes, and also revisiting the same location results in a new visited place. Visited places are then grouped into \textbf{(4)~journey fragments} as space-time clusters of temporally consecutive visited places within a shared geographic region.
Finally, the journey fragments are temporally ordered to form the global \textbf{(5)~journey path}, linking all images of the archive into a continuous trajectory.

Each journey fragment is visually represented as a miniature map, for which we use the \emph{transverse} Mercator projection~\cite{snyder1983map} with a flexible projection center, delivering unique perspectives on the world with low cartographic distortion.
The zoom level, spatial extent, and projection center are dynamically determined based on the spatial distribution of the locations and their centroid within each journey fragment, ensuring that all places are visible and that the available space is used effectively. 
Within each map, visited places are rendered as colored dots, connected by a smooth interpolated curve in temporal order. Both dot size and curve width scale with the number of photos. The curves are drawn using Catmull-Rom splines, which create smooth paths that resemble meandering rivers and pass directly through the visited places. 

\begin{figure}[t]
  \centering
  \includegraphics[width=\columnwidth, alt={Two-panel figure. Panel A labels the main elements of the meandering path arrangement: A vertical timeline, the journey path and its fragments, a circular miniature map follower on the journey path, and a horizontal stream of images and keywords coming out of the journey path. Panel B shows four stages of a transition in which circular journey fragments move from the aligned grid through progressive rearrangements into a vertical meandering path. The fragment maps gradually disappear, their individual paths connect to form one continuous journey path, and a miniature map follower appears.}]{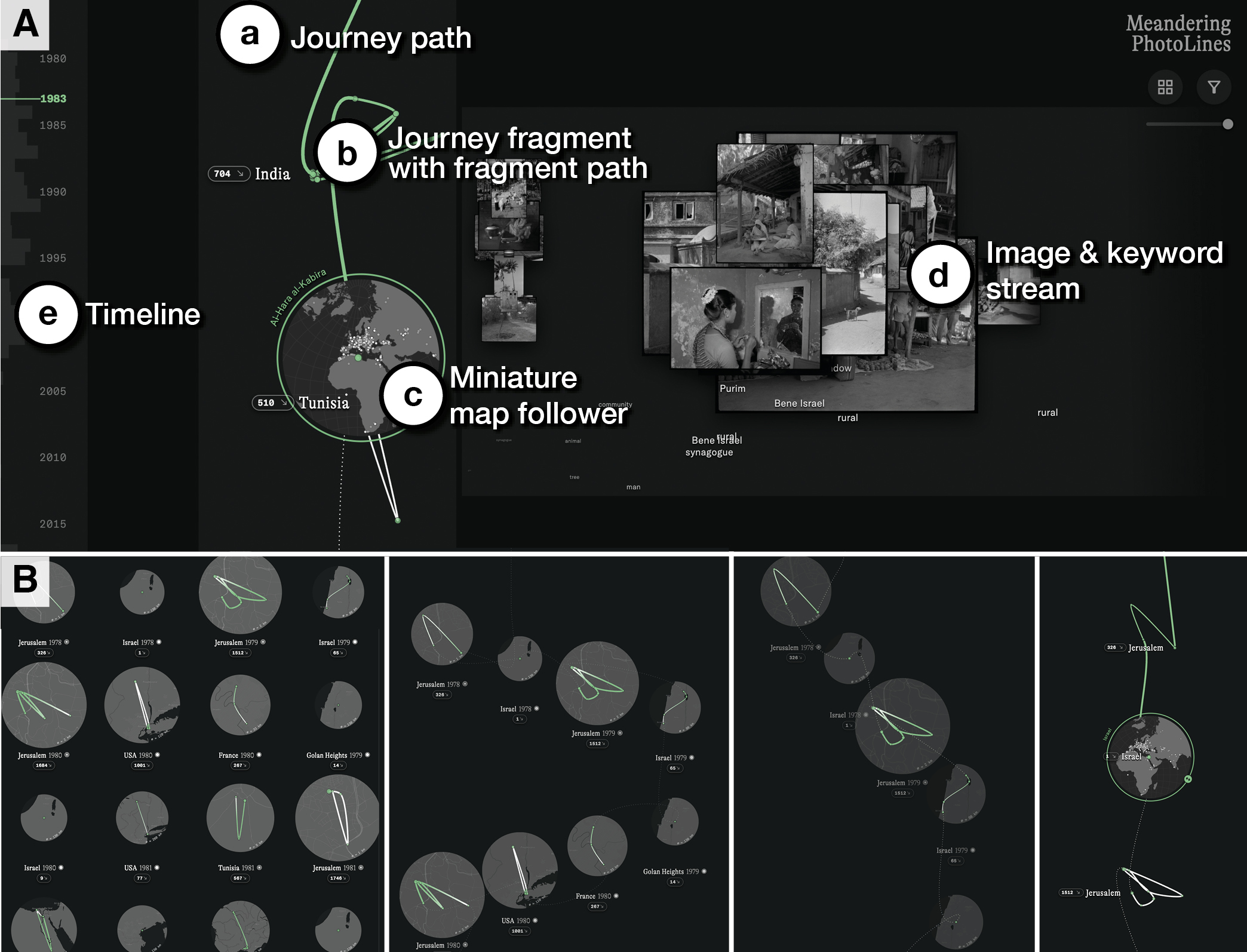}
  \caption{(A) Overview of the interface elements of the meandering path. (B) Fluid transition between the aligned grid and the meandering path.}
  \label{fig:transition}
\end{figure}

\subsection{Meandering Path}

The meandering path (\autoref{fig:transition}A and \autoref{fig:teaser}E) presents the journey path as a continuous, vertically oriented curve (\autoref{fig:transition}A-a) 
connecting all fragment paths (\autoref{fig:transition}A-b) top to bottom while preserving the local orientation of visited places. 
A \textbf{miniature map follower} (\autoref{fig:transition}A-c) is superimposed on the vertical curve. It follows the currently focused visited place and provides spatial context. Its relation to a vertical \textbf{timeline} (\autoref{fig:transition}A-e) indicates the temporal context. The timeline also indicates the time span of the archive, and histogram bars show the number of photos per year.
Unlike many cartographic approaches that rectify routes into simplified vertical paths, our approach does not morph the journey fragments themselves, but vertically repositions and connects the fragments. Location labels mark the beginning of each journey fragment and the number of contained images.
The photos and keywords associated with the currently focused visited place are dynamically revealed left-to-right in an \textbf{image \& keyword stream} (\autoref{fig:transition}A-d) next to the miniature map follower (DG1). 

The central mode of interaction is \textbf{vertical scrolling}. It allows users to travel from one visited place to the other, akin to meandering along a journey over time. When a visited place is approached by scrolling, the miniature map follower automatically re-centers its cartographic projection, offering an unbiased, low-distortion depiction of the spatial context. At the same time, preview images and keywords associated with the newly visited place emerge from the image \& keyword stream, starting as small thumbnails that grow and drift to the right as scrolling continues, gradually leaving the screen.
Scrolling back reverses the movement, with images and keywords gradually disappearing from the stream. 
This appearance and disappearance mimics the temporal experience of following a journey: Not everything is visible at once, and what comes next remains unknown, aiming to create moments of surprise and discovery (DG3). 
In addition to vertical scrolling, the timeline also supports jumping between years by dragging or selecting a year label. To balance the visibility of the spatial context, the miniature map follower can be resized by dragging its outer border.

To reveal the complete photographic material within a journey fragment and make its serial character accessible (DG1, DG2), users can click on the miniature map follower or a path label (\autoref{fig:transition}A-c). Doing so \textbf{unfolds a fragment path} into a straight line (\autoref{fig:teaser}F), similar to \textit{Tied in Knots}~\cite{elli2020tied}.
This unfolding reveals all image series along the fragment, including their associated locations and dates. Following a piling approach to foreshadow content~\cite{lekschas2020generic}, the images of a series are aggregated into piles, where the number of images is represented through the size of a visual pile. Hover interaction fans out the images to preview the full extent of a series. Clicking on a series opens a detail view (\autoref{fig:teaser}G), revealing metadata (e.g., title, location, date, keywords) and zoomable individual images. A close button allows users to fold the images back into their compact path representation (DG2).

\subsection{Aligned Grid}

The aligned grid (\autoref{fig:teaser}A–D) shows all journey fragments as circular \textbf{miniature maps} in a \textbf{small-multiples layout}, effectively creating a spatiotemporal overview. The maps are ordered temporally and arranged in a zigzag pattern (\autoref{fig:concept}). 
The direction of time within each journey fragment is indicated through a white-to-green color gradient along the fragment path. Depending on the spatial extent of the journey fragments, the maps show different levels of detail in terms of natural (e.g., rivers) and cultural (e.g., urban areas, city labels) features. A label indicates the scale of each map, which can range from local ($\approx 2~ km$ diameter) to continental scale ($\approx 4000~km$). 
The size of each map reflects the number of images of the journey fragment. 
Further details can be revealed interactively. Hovering a visited place reveals a corresponding label and related image previews. Hovering the map background triggers an in-situ slideshow of the journey fragment's photographs (\autoref{fig:teaser}C). Alternatively, depending on whether users prefer spatial context or photographic content (DG1), 
the miniature maps can be replaced globally with photo previews (\autoref{fig:teaser}D).

\begin{figure}[t]
  \centering
  \includegraphics[width=\columnwidth, alt={ Two photographs of Meandering Photo Lines on differently sized touch displays. Panel A shows a smartphone held and used with one hand, with the journey arranged vertically on its screen. Panel B shows a person interacting with a large vertical touch display, where the same visualization fills the exhibition-scale screen.}]{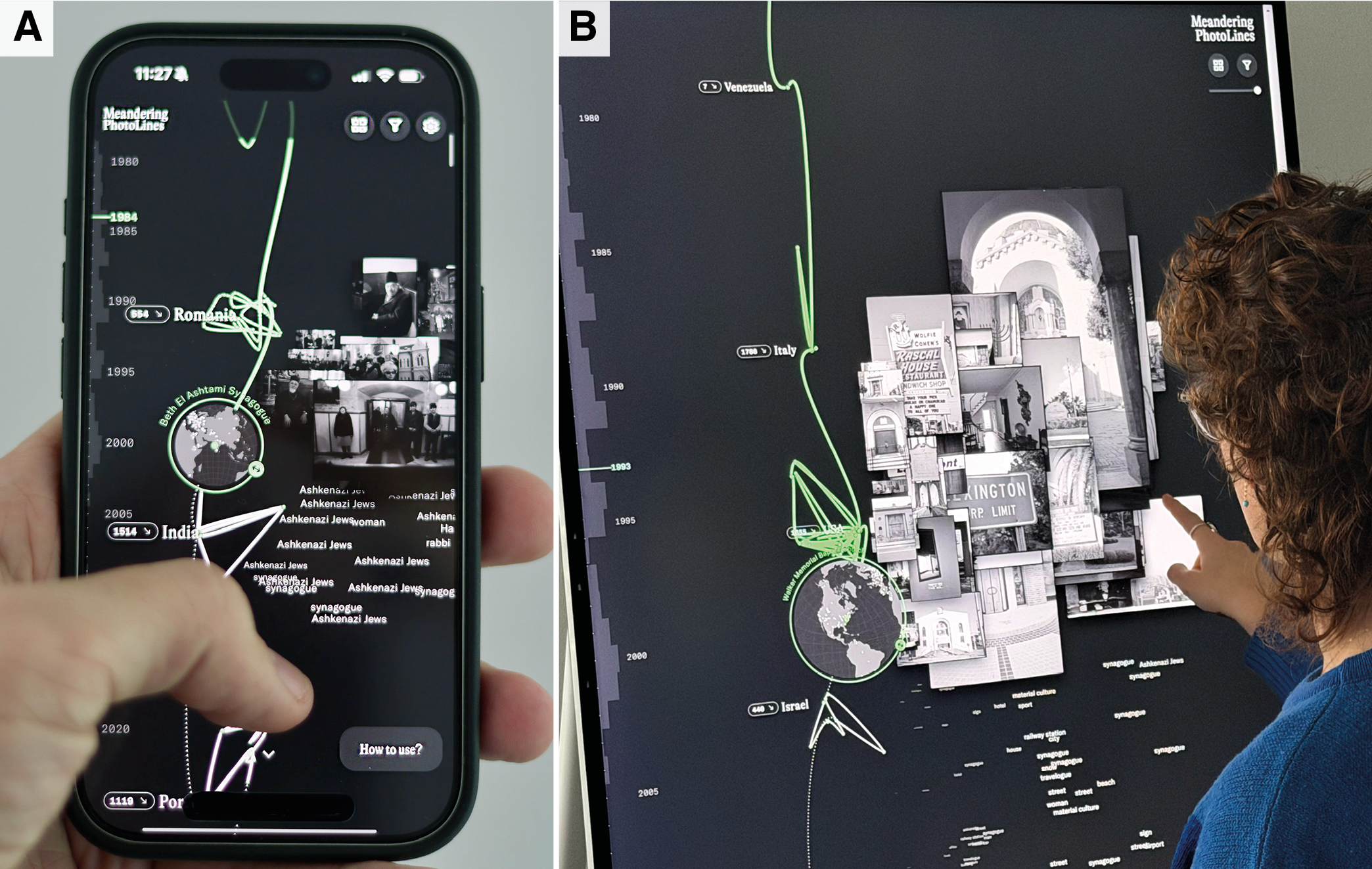}
  \caption{Our visualization adapts to different screen sizes, from (A) small smartphone screens to (B) large touch displays (DG4).}
  \label{fig:responsivedesign}
\end{figure}

\begin{figure*}[t]
  \centering
  \includegraphics[width=\textwidth, alt={Four screenshots of the same personal photo collection at increasing levels of detail. From left to right, they show a grid of abstract journey glyphs labeled by country, a grid of circular map tiles with their fragment paths, the meandering path layout with a map follower and the image and keyword stream, and an unfolded image series containing desert photographs.}]{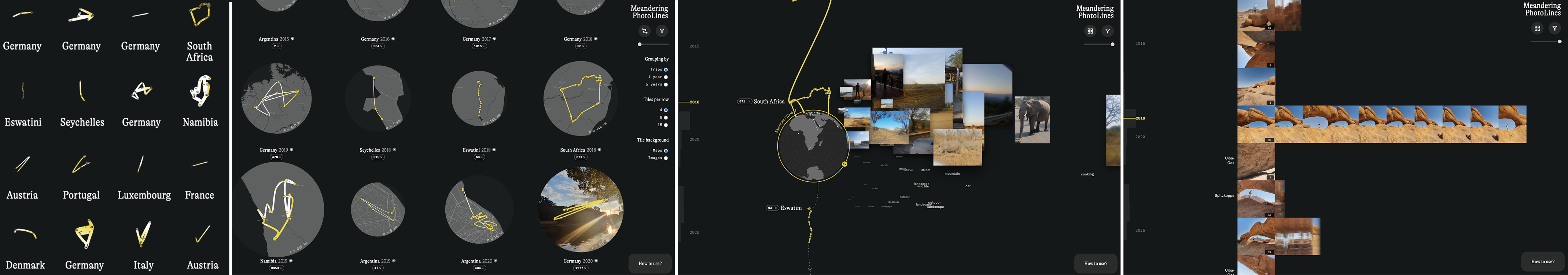}
  \caption{Selected views from the second case study illustrating the application of our approach across multiple scales of a personal photo collection.}
  \label{fig:casestudies}
\end{figure*}

Similar to \textit{Visits}~\cite{thudt2013visits}, different clustering strategies can be used to adjust the granularity of journey fragments (DG2). The default setting, referred to as ``trips'' (\autoref{fig:teaser}C), combines a geopolitical unit (e.g., country level) with a temporal constraint (e.g., within one year). 
In contrast to the distance-based metrics used in \textit{Visits}, our clustering relies on semantic geographic units (e.g., country), which results in more recognizable location labels and better aligns with common mental models of visited places.
Alternatively, purely temporal clustering without geographic constraints can be used  to create a temporally structured global overview of visited places, such as one-year or five-year groupings of the photographs (\autoref{fig:teaser}B).

The aligned grid can further be adapted in terms of the number of tiles per row. Making more tiles visible provides a broader overview and improves comparability between fragment paths~(DG3). To keep small or dense tiles legible, the tiles responsively reduce their semantic and visual details as available display space decreases.
At the lowest level of detail (many tiles per row), only fragment labels and paths are shown, providing an overview of the different movement patterns in form of individual \textbf{signatures} across the photographic archive (\autoref{fig:teaser}A).

\subsection{Interaction Mechanisms}

To support joyful and coherent experiences \cite{bludau2025fluidly, elmqvist2011fluid}, interactions across both arrangements enable fluid navigation between scales (DG2) while remaining operable from small smartphone screens to large interactive installations (DG4, \autoref{fig:responsivedesign}). As described above, the grid relies on hover and click, while the path is explored by scrolling, timeline navigation, and on-demand unfolding. On large interactive displays in exhibition settings, the grid can serve as a shared overview inviting collective orientation, while the path supports more individual engagement (DG3).

Animated transitions (\autoref{fig:transition}B) between the two arrangements---controllable via both a button and a slider in the top right---are included to connect their interaction modes into a coherent exploration experience (DG2). While click-triggered transitions are simple to use, the slider allows users to manually control the transition speed, supporting a gradual understanding of the transformation~\cite{bludau2025fluidly}. Additionally, both arrangements are not only connected through a global-view change button, but also selecting a tile or an individual place in the aligned grid transforms the layout into the meandering path, automatically scrolling to the corresponding position and unfolding the path to directly reveal the images within their spatiotemporal context (DG1, DG2).

A filter \& search panel provides a searchable list of all keywords in the archive, ordered by frequency, enabling serendipitous meandering along keyword-specific paths. When a keyword is selected, the entire photographic journey is filtered to include only photos, photo series, visited places, and journey fragments associated with the selected keyword. The aligned grid then shows only the journey fragments that contain matching photos (\autoref{fig:teaser}D). 
The meandering path is re-constructed, and the image \& keyword stream and the unfolded photo series are filtered accordingly. This filtering effectively creates new journeys through the archive, for example, focusing only on landscape or architectural photographs. This can reveal patterns of topics across time and space, enabling an overview of evolving photographic motifs.

\subsection{Technical Implementation \& Visual Language}

We implemented Meandering Photo Lines as a static website using web technologies and the data visualization library D3.js~\cite{bostock2011d3}. 
To ensure long-term availability, the prototype operates without database queries or server-side components, using pre-processed static CSV and JSON files that are loaded incrementally on-demand when entering a hierarchy level. For the miniature maps, we use vector data from \href{https://naturalearthdata.com}{Natural Earth}. 
For details in highly zoomed-in areas, we additionally use street data from OpenStreetMap, retrieved via the \href{https://overpass-api.de}{Overpass API} during pre-processing rather than at runtime, using conservative timeout settings and minimal bounding-box extraction. To improve performance and minimize live rendering, the fragment base maps are pre-rendered as image files in a secondary script. Scroll-based interactions rely on the native JavaScript Intersection Observer. On touch devices, hover interactions are mapped to a first tap while a second tap triggers the click action. 
To support interlinking within storytelling formats~\cite{eingartner2025inflecting} and bookmarking of specific views, interaction states are encoded in the URL hash. 
A configurable Python pipeline will consolidate the pre-processing into a ready-to-use data package for the visualization.

The visual design foregrounds the photographic material through a very dark gray background and a single configurable highlight color~(e.g., green), allowing deep blacks and colors within the photos to stand out.
A navigation bar on the right provides access to filtering, view switching, and layout adjustments. Icons and hover transformations aim to indicate interactivity throughout the interface, while an on-demand ``How to read'' section offers a short introduction.

%% file: sections/05_usecases.tex
\section{Case Studies}\label{casestudies} 
We examine the applicability of our approach through two case studies that differ in archive type, scale, and data characteristics. The first case study involves the previously mentioned professional photographer’s archive, which informed the design process. The second follows an autobiographical design approach~\cite{neustaedter2012autobiographical} by visualizing the personal smartphone photo collection of one of the authors.

\subsection{Professional Photographer's Archive}

This case study is based on the collaboration around the photographic archive of \textit{Frédéric Brenner}~\cite{bruggemann2025granularities}.
With ca. 100,000 images spanning roughly 1,000 places and 700 keywords, the archive documents Jewish diaspora from 1978 to 2021 in over 40 countries (\autoref{fig:teaser}).
Beyond documenting Jewish life, the archive traces the photographer’s own life, reflecting his artistic development from early documentary work, through large-scale productions, to more intimate portrait photography.

The archive has recently been digitized and manually cataloged, including metadata such as date, title, location, and keywords describing motifs and Jewish practices. Images were additionally grouped into series of related photographs, with selected highlights and published images marked as part of a curated edition, which in our approach are prioritized for previews. Due to the manual cataloging of the mostly analog material, the date of creation is mostly limited to the level of a year. However, the original ordering is largely preserved through numbered original image slides. The detail level of depicted places varies in granularity, but locations are linked to Wikidata URIs, providing geographic coordinates usable for visualization.
The goal of the visualization is not only to provide a web-based interface for exploration and analysis, but also to support use in exhibition contexts.

A distinctive characteristic of the archive is its detailed keyword cataloging, which enables the construction of topic-based journeys. For example, selecting keywords such as Purim (a Jewish holiday) or synagogue constructs paths composed only of corresponding images throughout the archive, revealing cultural and temporal variations in religious practices across different places. The grid layout shows broader patterns: phases of very high global mobility between 1980 and 2000 give way to a stronger focus on Israel and Germany after 2000, reflecting a shift in the photographer's geographic scope. Switching the grid background to images can for example visualize shifts in photographic style, from analog black-and-white to digital color. 
In the meandering path, the curated highlights are surfaced through the image stream, showcasing many of the photographer's most important works while still offering access to all other photos along the way.

\subsection{Personal Smartphone Photo Collection}

The second case study applies our approach to personal smartphone collections (\autoref{fig:casestudies}). As personal photo collections are typically oriented toward reminiscing or sharing with friends and family~\cite{thudt2015visual}, our visualization serves the purpose of reflection rather than analytics.

The personal archive comprises approximately 20,000 photographs across 280 places over 10 years. In contrast to the manually curated professional archive, this dataset is based on automatically extracted EXIF metadata (timestamps and GPS coordinates) and computationally generated groupings and keywords. 
Images were grouped into series based on temporal and spatial proximity as well as visual similarity using the \textit{ImageHash}~\cite{buchner2016imagehash} library, while keywords were assigned using the vision-language model \textit{CLIP}~\cite{clip2021} with a predefined vocabulary (e.g., portrait). Following recommendations for handling personal data~\cite{thudt2015visual}, location data was spatially aggregated prior to reverse geo-coding to reduce data density and protect privacy.

Overall, the visualization highlights patterns of work travels and vacations, and unlike conventional smartphone photo browsers, reveals movement patterns that would otherwise remain invisible. Representing photo series allows us to observe personal photographic behavior, for example, which subjects tend to be captured as individual snapshots (in our case landscapes) and which appear more often in series (e.g., people). The keyword filter enables the construction of thematic journeys, for example, by following paths composed only of animals.

\subsection{Observations and Reflections}

The two case studies demonstrate the applicability of our approach across different archive types and sizes, while also revealing overarching challenges related to photo archive data and differences between manually curated and automatically collected datasets.

\paragraph{Analog vs.\ digital photography.} The professional archive, largely based on analog photography and curated metadata, reflects typical challenges of humanistic data~\cite{drucker2011humanities}, including varying levels of detail as well as incomplete, uncertain, or inconsistent information. At the same time, its manually assigned keywords provide rich nuance, particularly for culturally specific aspects (e.g., Jewish holidays). However, temporal data is often limited to yearly granularity, and most locations are only specified at the city level.

In contrast, the predominantly digital photography within the personal archive offers highly precise temporal and spatial metadata. Here, challenges shift toward data privacy, as detailed location traces may expose movement patterns or personal information~\cite{thudt2015visual}. Consequently, reducing spatial precision becomes an important consideration. Moreover, personal photo collections are typically not manually cataloged, as they emerge from everyday life rather than curated photography. As a result, content exploration relies more on the quality of automated processing, for example, through computational methods or large language models. While basic keyword extraction performs well, capturing more nuanced or culturally specific content remains challenging.

\paragraph{Closed and living archives.}
Differences also emerge depending on whether an archive is closed (i.e., does not change over time) or living (i.e., continues to grow). 
From an auto-ethnographic perspective, the first author, as owner of the personal collection, observed that the existence of the technique influenced photographic behavior, encouraging more deliberate collection of locations through photography (``I don’t have this location yet'') and increased photo-taking to reconstruct movement. This reflexive effect, where the analytical tool shapes the practice it represents, points to broader dynamics relevant to personal data visualization. Subsequently viewing one’s own collection in this form also introduces a stronger emotional dimension compared to a more distant appreciation of another photographer's journey, as images are tied to memories of places, people, and personal experiences, while revealing previously unnoticed movement patterns.

\paragraph{Journey trajectories.}
Based on the two case studies, we identified five recurring journey patterns, visible particularly in the aligned grid's most zoomed-out view, where fragment paths are reduced to compact signature glyphs. Although derived from photo archives and not actual movement data, they can still help us better understand the photographic journeys. 
We can characterize phases in a photographer's career, compare mobility across archives, and---particularly for manually cataloged data---flag clusters where imprecise or missing data may have collapsed complex movement into overly simple shapes.

\smallskip 
\begin{wrapfigure}[5]{r}{0.07\textwidth}
\vspace{-8pt}
\hspace{-6pt}
    \includegraphics[width=\linewidth,
        alt={Schematic example of a line trajectory.}]{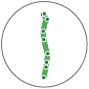} 
\end{wrapfigure}
\textbf{Line} trajectories form a sequential, line-like structure, with a defined start and end point, connected by intermediate locations, without significant detours or returns to previously visited places. Such trajectories may result from one-way-route traveling. 

\begin{wrapfigure}[5]{r}{0.07\textwidth}
\vspace{-12pt}
\hspace{-6pt}
    \includegraphics[width=\linewidth,
        alt={Schematic example of a round-trip trajectory.}]{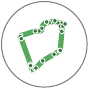} 
\end{wrapfigure}
\textbf{Roundtrip} trajectories share the same start and end point, with intermediate locations visited in a circular movement. Typical examples include multi-stop travel itineraries departing and returning to the same city, for example, a common airport.

\begin{wrapfigure}[5]{r}{0.07\textwidth}
\vspace{-12pt}
\hspace{-6pt}
    \includegraphics[width=\linewidth,
        alt={Schematic example of a hairball trajectory.}]{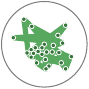} 
\end{wrapfigure}
\textbf{Hairball} trajectories result from high activity in many directions, producing densely overlapping and seemingly unstructured paths. These arise from high mobility within a small geographic radius or from aggregating long time frames into a single cluster.

\begin{wrapfigure}[5]{r}{0.07\textwidth}
\vspace{-12pt}
\hspace{-6pt}
    \includegraphics[width=\linewidth,
        alt={Schematic example of a starburst trajectory.}]{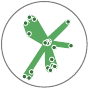} 
\end{wrapfigure}
\textbf{Starburst} shapes originate from one or more central locations, extending outward with repeated returns. This pattern indicates high mobility combined with a stable central base, such as city-based travel from a fixed accommodation or photographic activity in a home location.

\begin{wrapfigure}[4]{r}{0.07\textwidth}
\vspace{-12pt}
\hspace{-6pt}
    \includegraphics[width=\linewidth,
        alt={Schematic example of a single-point trajectory.}]{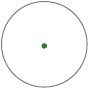} 
\end{wrapfigure}
\textbf{Single points} result from low photographic activity or low mobility within a spatiotemporal cluster. For cataloged data, single points associated with many images can also point to imprecise or missing location data.

\smallskip 
These patterns only capture a few of the most recognizable structures and do not aim for completeness. Many journeys are hybrid in nature. 

%% file: sections/06_evaluation.tex
\section{Qualitative feedback}\label{evaluation} 

While the design process was closely tied to iterative collaboration with a photo archive, we gathered additional qualitative feedback through a think-aloud study~\cite{Carpendale2008} with eight participants (P1--P8).
As our approach has no functional equivalent to serve as an adequate baseline, we use this exploratory format, focusing on how participants experience and make sense of the visualization and the different parts of the interface.

\subsection{Setup and Procedure}

Due to its broader relevance for the general public, we used the professional photo archive for the think-aloud study. Participants were recruited through two semi-public calls and targeted outreach around the archive, resulting in eight participants aged 20--60 (four below and four above 40; five female, three male).  
Participants were selected based on a background or interest in photography, familiarity with the archive, or domain knowledge related to the Jewish diaspora. Sessions lasted 30--45 minutes and were conducted remotely via Zoom with screen sharing, using participants' own devices. 
After a short introduction to the study context and the archive, participants were sent a link to the prototype and asked to freely explore it while thinking aloud about what they were doing, thinking, and interpreting. 
The prototype functions were not introduced beforehand. Instead, on-boarding popups explained the main features, which participants could read or skip. To balance initial entry points, half of the participants started in the aligned grid and the other half in the meandering path. 
The study followed the ethics regulations of the authors' institutions, and participants provided informed consent for recording, transcription, and analysis.

\subsection{Observations and Findings}

Based on transcriptions of the sessions, statements were inductively coded into thematic groups, ranging from general usability observations to aspects specific to our approach. Below, we summarize our findings.

\paragraph{\textbf{Support of engaging exploration.}}
Participants consistently described the interface as engaging and visually appealing: \textit{``I love the graphics. It’s beautiful in a way''}~(P8).
\textit{``That’s of course really cool how this globe here is now rotating. That’s really fun to interact with''}~(P1).
The visual structure of the map-path combination was described several times as an interesting and aesthetic structure: \textit{``I find these maps here very nice, somehow they catch me''}~(P7). \textit{``I like the lines that are in the circles. That's kind of interesting, also, like, visually''}~(P2).
While the visual design of the maps was widely perceived as aesthetically appealing and understood as the journey, the directionality of the path was less clear within the grid, and one participant noted that certain visual elements evoked slight unintended associations in the context of current global events, e.g., \textit{``drone flights’’}~(P5).

Some participants noted how they \textit{``get pulled in''}~(P5) during exploration: \textit{``You can definitely get lost in here. That's wonderful''}~(P8).
The fanning out mechanism of the image series in the path view was highlighted by all participants, describing it as engaging, beautiful, or helpful: 
\textit{``I definitely find that super cool. This peek and pop, that I hover on top and then all the other images somehow pop out. I think that’s a very, organic interaction once you’ve understood it, and it definitely motivates me as well. So I'd now want to look at all of them, simply because it’s somehow interesting to see what’s behind it''}~(P7).

Finally, our observations suggest that our approach can support one of the key purposes of photo archives by enabling participants to engage with and appreciate the photos:
\textit{``These are already really nice photos. Good portraits''}~(P6).
\textit{``Super great to be able to take in all these different life realities through the images like this''}~(P1).

\paragraph{\textbf{Between meandering and analysis.}}
Participant feedback further suggests that our visualization supported understanding of the photographic journey: \textit{``I actually found it very interesting that you can see: okay, within that short period of time, somehow different journeys were undertaken into different areas, and to actually be able to trace his path through these lines''}~(P7).
The aligned grid was mainly perceived as a clearer overview supporting more structured, goal-oriented exploration, with one participant highlighting the grouping by years and trips as helpful for quickly gaining a sense of scale and distribution: \textit{``I get a sense of the amount very quickly. Especially being able to group things by years and trips like this. I find that super helpful, because I immediately get a feeling for where and how much''}~(P5). In contrast, the meandering path was experienced as more immersive and engaging, encouraging curiosity-driven browsing: \textit{``it has something like Indiana Jones, where you’re flying to a journey of discovery''}~(P5).

Some participants preferred the more conventional structure of the grid: \textit{``So, I find the tile view gives me a quicker overview. […] I find I reach my goal faster that way''}~(P3). Others were more drawn to qualities of the path: \textit{``What I find really great here is that I can immediately see the images in the context of their location. […] This view should be the primary one, simply because it is much more compelling''}~(P1). 
\textit{``I can continue to surf further in this topic in this way, and I don’t even have to go to the individual places, but can just scroll like this. […] I find it very nice how the photos pop up like that and countries and years move along''}~(P3).

Despite clear, and sometimes strong, personal preferences for the aligned grid (P3, P4) or the meandering path (P1, P2, P6), both were described as serving complementary purposes: \textit{``So if you want to understand the journey better, you need the line, and if you want to understand the thing as a whole, you need the grid''}~(P2). \textit{``But whatever you want, whether just kind of meandering through the images or doing a really intense analysis and looking for something very specific, like for example a clay oven. It can give you both''}~(P5).

With regard to understanding photographic processes, participants appreciated the access to all images: \textit{``I also find it great that you can see all the photos. Because some are a bit underexposed. And this one here is probably the better photo, but I think especially in comparison it shows me like, nice: The guy also just experimented, just like I would do. That makes the whole thing more relatable for me and also has a certain kind of enjoyment in it''}~(P5).
In contrast, one participant preferred fewer, more curated images, stating they would \textit{``rather have fewer images but better images''}~(P6).

Notably, nearly all participants expressed a desire for additional contextual information about the photographer and his journey---pointing not to the visualization of the journey itself but to the content it displayed: \textit{``So, I think what has definitely become much clearer to me is how the photographer moves within a region, and that is also super interesting. But of course that raises the question of why he moved the way he did. And that’s something that isn’t answered here''}~(P1).

\paragraph{\textbf{Learning by doing.}}
Only two participants (P2, P8) used the offered on-boarding assistance of the prototype.
Several participants stated they would normally not engage with an on-boarding and prefer to explore functionalities themselves: 
\textit{``I would actually click it away and hope that I somehow sort it out myself''}~(P7).
Accordingly, many visual representations or interactions were not directly clear, and participants could be actively observed making sense of the interface through interaction.  Animated transitions and unfolding structures were perceived as supporting this gradual understanding as participants moved between arrangements: \textit{``Okay, it unfolds, and the further I move to the right, it unfolds downward''}~(P5). Another participant described how the transitions after a selection in the grid \textit{``give the feeling that you are actually going somewhere''}~(P8).
The fluid transitions appeared to aid orientation between overview and detail rather than adding confusion.

The arrangement and selection of images within the image stream of the path was not immediately legible to all:
\textit{``I’m trying to figure out, when I scroll through, whether images with keywords somehow pop up on the right automatically, but it’s not yet entirely clear to me where they come from or why this image appears now and not some other one''}~(P7). 
Typical patterns to increase understanding included slow up-and-down scrolling in the path or verification of assumptions by looking into the detail views:
\textit{``I'm thinking about how this globe corresponds with the images next to it. For instance, you see it's the Czech Republic. And this makes sense, because this is the cemetery in Prague. Okay, now I see. They come from it. Oh, they returned to their location. Oh, that's nice, I get it now''}~(P8).  
However, participants noted that they were willing to accept exploratory learning-by-doing: \textit{``I had the feeling that I had to engage with it a bit to be able to use it. But it’s also so interesting that you also feel like doing that''}~(P4).

\paragraph{\textbf{Additional usability insights.}}
Modes of interaction varied among participants. Some primarily navigated via the selection of place-time clusters in the aligned grid (P3, P4), while others relied more on scrolling and timeline-based navigation within the meandering path (P6, P7), or mixed approaches. For targeted exploration of topics, participants consistently understood and successfully used the filter functionality to explore thematic journeys across the archive:
\textit{``Okay, now I see all the different journeys that contain images that are probably tagged with the keyword ‘Family’. And now I want to go back to the timeline and that works, great''}~(P7). Some participants requested extended search/filter capabilities (e.g., people search)~(P1, P3, P5, P8).

Participants also expected certain interface conventions that were not yet present, including support for the browser back button (P5), options to download images (P5), horizontal scrolling within the image \& keyword stream (P4, P5, P8), and the ability to directly select visited places via the miniature map follower (P3, P4). In this context, the resizing of the map follower through dragging did not work well, with several participants expressing a preference for a simple click-based interaction (P1, P5, P7).
For accessibility, participants highlighted the need to adjust text sizes and to provide simplified explanations in the on-boarding (P4, P5). More generally, we observed that some buttons and interactive elements could benefit from a more descriptive labeling.

%% file: sections/07_discussion.tex
\section{Discussion}\label{discussion} 

Next, we reflect on the findings and implications for our approach.

\paragraph{Exploration patterns.}
The qualitative observations suggest that the approach can support the presentation of photographic material within its spatiotemporal context~(DG1), while enabling both open-ended and focused forms of exploration. 
Clear differences in preferences became apparent: Some participants gravitated toward the structured overview of the grid, others toward the immersive quality of the meandering path, yet both groups described the two views as complementary. This suggests that supporting multiple modes of engagement within a single interface is not redundant but necessary to enable analytical (e.g., understanding topic distributions) and curiosity-driven exploration (e.g., following along and selecting interesting photos)~(DG3). While the animated unfolding of the multiscale navigation appeared to aid comprehension across scales~(DG2), not all encoding and interaction mechanisms were understood at a glance.  At the same time, the willingness of participants to engage with unfamiliar representations raises the question of whether unconventional visualization mechanics may even enrich the experience in certain usage contexts.

\paragraph{Narrative context.} 
Most participants expressed a desire for more narrative context about the photographer and his journey. In favor of generalizability and transferability to personal collections, we consciously did not integrate narrative content directly. However, the future archive platform, where Meandering Photo Lines will be used, will have interlinked experiences focusing on narrative aspects. More broadly, we see a need for further research into bridging data visualization and narrative dimensions in the context of photographic archives.

\paragraph{Data quality.} 
Consistent archiving practices and the knowledge of the living photographer, combined with detailed diaries, provided temporal and spatial metadata for nearly all images of the first case study, although the granularity of locations and dates varies, and correct temporal order depends on the accuracy of the archived film rolls, leading to individual inaccuracies. These are typical challenges of cultural heritage data, but despite such uncertainties, our approach can still offer useful insights and help identify data issues. While the visualization can accommodate varying levels of spatiotemporal granularity, the quality and precision of the resulting representations depend on the available metadata, as spatial and temporal relationships form the core structure of the visualization.
For the location clustering, we relied on geopolitical boundaries due to the inherent archival organization around country-level journeys within the professional photo archive, while following Library of Congress guidelines regarding geopolitical entities. This poses several challenges, as borders change over time or are internationally disputed. We consciously decided against displaying country borders within the maps. Depending on the collection and usage context, alternative or manually adjustable clustering approaches, similar to \textit{Visits}~\cite{thudt2013visits}, may also be appropriate.

\paragraph{Scalability.}
The two case studies demonstrate the approach's applicability to large-scale photo archives, including collections containing around 100,000 photographs. The scalability is supported by the multiscale and hierarchical organization of the visualization and animated transitions between states (DG2), reducing visual complexity and the need to access and render all image elements simultaneously while also enabling flexible use across display sizes (DG4).

\paragraph{Limited evaluation.}
While related systems such as smartphone photo galleries exist, a direct comparison is not meaningful, as our approach integrates photo browsing with spatiotemporal journey exploration in a way that makes it difficult to isolate individual features against an existing baseline. Instead, we followed a qualitative approach informed by the earlier workshops, focusing on how users experience and engage with Meandering Photo Lines rather than on performance metrics. While the study does not compare against other techniques, it still allows comparisons between views within Meandering Photo Lines, showing, for example, that the preferred arrangement depends on the individual user.
Our qualitative findings are based on eight participants with prior interest in the archive or photography, exploring remotely on their own devices. We did not evaluate the personal smartphone case study with external participants, nor did we observe usage in exhibition settings or on large displays, where interaction patterns may differ significantly.
Further evaluation is needed to better understand the effectiveness of our approach, particularly given the diversity of preferences observed. This may include more quantitative evaluation of interaction patterns, usage observations in exhibition settings, and the application of the approach to other collections.

\paragraph{Time, space, and photography.}
Balancing the dimensions of time and space and the visual qualities of the photographs was a central objective. 
The variants we explored can be read as positions along the axes of a design space: spatiotemporal emphasis vs. direct photo integration, grid vs. path arrangement, and granularity of clustering.
The journey signatures in the aligned grid abstract a set of photographs into a single spatiotemporal glyph while keeping the underlying images only one interaction away.
At the opposite end, photographs are directly embedded along the meandering path with little abstraction, foregrounding the images themselves.
The spatiotemporal structure aligns more closely with classical space-time approaches, while more direct photo integration poses challenges due to the large amount of image data and the implications of embedding photographs along continuous paths. We believe that our implementation provides a scalable balance between spatiotemporal structure and photographic detail. At the same time, the explored variations suggest a broader design space around the integration of photographic material and spatiotemporal representations that may provide interesting directions for future research.

%% file: sections/08_conclusion.tex
\section{Conclusion}\label{conclusion} 
Photographic archives carry within them the traces of journeys. With \textit{Meandering Photo Lines}, we have explored what happens when these traces are given a visual form of their own, rather than remaining buried in metadata. The resulting spatiotemporal pathways, navigable from distant signature glyphs down to individual photographs, offer a mode of engagement that emphasizes the continuity of the photographic journey and the serial character of photographic work. Conventional photo archive browsers organize collections through thumbnail grids, but leave the spatiotemporal trajectory without a visual form of its own. The movement patterns, their comparison across journeys, and the contained topics remain inaccessible in these tools.

With our work, we have shown that following a meandering path through a collection can deepen understanding of photographic practices.
Our qualitative observations suggest that our approach can support both analytical examination and curiosity-driven exploration, with participants describing the two complementary views as serving distinct purposes. 
At the same time, we identified several directions for future work.
Besides storytelling integration, further evaluation, and a broader exploration of the emerging design space, we envision extensions such as animated journey playback for exhibition settings, printable journey overviews as collection portraits, and the integration of video material.